\documentclass[preprint, onecolumn]{revtex4}
\usepackage{graphicx}
\usepackage{physics}
\usepackage{chngpage}

\def\bse{\begin{eqnarray*}}
\def\ese{\end{eqnarray*}}
\def\be{\begin{eqnarray}}
\def\ee{\end{eqnarray}}
\def\bq{\begin{equation}}
\def\eq{\end{equation}}
\def\pr{\hbox{pr}}

\begin{document}
\begin{flushright}
MI-TH-1918
\end{flushright}
\title{Searching for Beyond the Standard Model Physics with COHERENT Energy and Timing Data}
\author{Bhaskar Dutta$^{\bf a}$}
\author{Shu Liao$^{\bf a}$}
\author{Samiran Sinha$^{\bf b}$}
\author{Louis E. Strigari$^{\bf a}$}
\affiliation{$^{\bf a}$ Mitchell Institute for Fundamental Physics and Astronomy,  Department of Physics and Astronomy, Texas A\&M University, College Station, TX 77843, USA}
\affiliation{$^{\bf b}$ Department of Statistics, Texas A\&M University, College Station, TX 77843, USA}
\begin{abstract}
We search for beyond the Standard Model (BSM) physics by combining COHERENT energy and timing data. Focusing on light, $\lesssim$ GeV mediators, we find the data favor a $\sim 10-1000$ MeV mediator as compared to the Standard Model (SM) best fit at the $\lesssim 2\sigma$ level. The best-fit coupling range is $g \sim 10^{-5}-10^{-3}$. The timing data provides statistical information on the neutrino flavor distributions that is not attainable from the nuclear recoil energies alone. This result accounts for uncertainty in the effective size of the neutron distribution, and highlights the power of including uncertainties on experimental backgrounds, nuclear structure, and BSM physics.
\end{abstract}
\maketitle
	
\section{Introduction}
The COHERENT collaboration has reported the first detection of coherent neutrino-nucleus elastic scattering (CE$\nu$NS)~\cite{Akimov:2017ade}. COHERENT utilizes the Spallation Neutrino Source (SNS) with a stopped-pion beam, which produces a well-known neutrino spectrum from pion and muon decay at rest. Muon neutrinos, $\nu_\mu$, thus arrive from a prompt decay of charged pions, whereas $\bar{\nu}_\mu$ and $\nu_e$ are produced from the delayed decay of muons. With an exposure of 14.6 kg-days, the COHERENT collaboration measured a best fit of $134 \pm 22$ nuclear recoil events from CE$\nu$NS, which is well in excess of the expected background events for this exposure. 

\par The COHERENT data provides an important new channel to search for beyond the Standard Model (BSM) physics. For example, the COHERENT data constrains non-standard neutrino interactions (NSI)~\cite{Ohlsson:2012kf,Miranda:2015dra} which may arise from heavy or light mediators~\cite{Coloma:2017egw,Coloma:2017ncl,Liao:2017uzy,Dent:2017mpr,Billard:2018jnl, Lindner:2016wff, Farzan:2018gtr, Brdar:2018qqj}. The data also provide novel constraints on new physics that manifests through the neutrino sector, including generalized scalar and vector neutrino interactions~\cite{AristizabalSierra:2018eqm}, as well as hidden sector models~\cite{Datta:2018xty}. It  also sets independent constraints on the effective neutron size distribution of CsI~\cite{Ciuffoli:2018qem,AristizabalSierra:2019zmy,Papoulias:2019lfi}, and on sterile neutrinos~\cite{Kosmas:2017zbh,Blanco:2019vyp}. Low mass mediator models are particularly interesting since there are no Large Hadron Collider (LHC) constraints for mediator masses $\lesssim$ GeV, and these models connect to new ideas associated with sub-GeV dark matter in the range MeV to GeV~\cite{Essig:2013lka}. The aforementioned constraints on BSM physics have mostly been extracted from the energy distribution of nuclear recoil events. 

\par The COHERENT data do not directly identify the flavor components of the neutrino flux, though it is possible to make an estimate for the contribution of the different flavors. For example the prompt $\nu_\mu$ component may be estimated from a timing cut, while the delayed components may be extracted from their spectral signatures. Previous authors have classified the events as prompt or delayed using a two-bin analysis in timing space~\cite{Denton:2018xmq}. Including timing information in this manner has the potential to strengthen constraints on NSI parameters, in comparison to those obtained from using information in the energy distribution alone. 

\par In this paper, we perform a likelihood analysis on the COHERENT data that utilizes the full energy and timing distributions of nuclear recoil events. These distributions provide information on the flavor components of the detected neutrino flux that is not available when considering the energy data alone, or when splitting the timing data into prompt and delayed events. We test for deviations from pure Standard Model (SM) interactions, considering as an example light mediators that couple to the SM.  We consider a scenario in which the quark-$Z^\prime$ coupling contains a momentum dependent factor associated with  hidden sector interactions. We show that including the timing distribution data adds substantial statistical constraining power, and in fact favors a BSM interpretation of the nuclear recoil data at the $\lesssim 2\sigma$ level. We also consider constraining the LMA-Dark solution using the timing data.

\section{Light mediators and Non-Standard Interactions}

\par In this section we briefly discuss the SM prediction for CE$\nu$NS. We introduce a model for light mediators as an example of BSM physics, and highlight how this model alters the predicted CE$\nu$NS cross section. 

\par In the SM, the differential cross section for a neutrino scattering off of
a target electron or quark of mass $m$ through a $Z$ exchange is
\begin{equation}
\frac{d\sigma}{dE} = \frac{G_F^2 m}{2\pi}\left((g_v+g_a)^2 +
(g_v-g_a)^2\left(1-\frac{E}{E_{\nu}}\right)^2 +
(g_a^2-g_v^2)\frac{m E}{E_\nu^2}\right)
\,,
\label{SM}
\end{equation}
where $G_F$ is the Fermi constant, $E$ is the recoil energy, $E_\nu$ is the incident neutrino energy,  $(g_v,g_a)= (T_3-2Q_{\rm em}{\rm{sin}}^2\theta_W,T_3)$ are the vector and axial-vector couplings to the $Z$-boson, $T_3$ is the third component of the weak isospin,
$Q_{\rm em}$ is the electromagnetic charge, and $\theta_W$ is the weak mixing angle ($T_{3} = -1/2$ in our convention). For quarks in the nucleus, if the momentum transfer between the neutrino and the nucleus is smaller than or comparable to the inverse size of the nucleus (typically $E \lesssim\mathcal{O}(50)$ MeV), coherent scattering occurs. For coherent elastic scattering off of nuclei, the axial vector contribution is proportional to the nucleus spin~\cite{Barranco:2005yy}, therefore it is negligible as compared to vector contribution.
The cross section for scattering on the nucleus is:
\begin{equation}
	\dv{\sigma}{E} = \frac{G_F^2Q_V^2}{2\pi}m_N\left(1-\left(\frac{m_NE}{E_{\nu}^2}\right) + \left(1-\frac{E}{E_{\nu}}\right)^2\right)F(q^2)
\end{equation}
where $F\left(q^2\right)$ is the nuclear form factor, $Q_V \equiv Z\left(2g^u_v+g^d_v\right)+N\left(g^u_v+2g^d_v\right)$, and $m_N$ is the nucleon mass. 

\par The form factor, $F(q^2)$, is a source of uncertainty in the scattering cross section, in particular at the highest nuclear recoil energy. Since there are no direct experimental measurements of the neutron distribution in the nucleus for CsI,  $F(q^2)$ may be estimated from the CE$\nu$NS data itself. Standard parameterizations are the Helm model~\cite{Lewin:1995rx}, or the symmetrized Fermi model~\cite{Piekarewicz:2016vbn}. Each of these model paramerizations are functions of the size of the neutron distribution, $R_n$, which can be viewed as a parameter in each model. For each of these form factor parameterizations and including only SM contributions to the scattering rate, the best-fitting neutron radius from the COHERENT data is $R_n \simeq 5.5$ fm~\cite{Ciuffoli:2018qem}.

\par The SM cross-section above may be modified due to a new mediating particle which couples to neutrinos and either electrons or quarks. For example, consider a new mediator $Z_{\mu}'$ with the following interaction terms:
\begin{equation}
\mathcal{L} \supset Z_{\mu}'(g_{\nu}'\bar{\nu}_L\gamma^{\mu}\nu_L + g_{f,v}'\bar{f}\gamma^{\mu}f + g_{f,a}'\bar{f}\gamma^{\mu}\gamma^5 f),
\label{Zprime}
\end{equation}
where $g_{\nu}'$, $g_{f,v}'$, and $g_{f,a}'$ are constants associated with new physics. With this interaction we can redefine the couplings $(g_v,g_a)$ of Eq.~(\ref{SM}) in the following way
\begin{equation}
(g_v,g_a) \,\Rightarrow\, ( g_v, g_a )
+ \frac{g_{\nu}'\, (g_{f,v}',\pm g_{f,a}')}{2\sqrt{2}G_F\, (q^2 + M_{Z'}^2)},
\label{eq:SMmod}
\end{equation}
where $q^2$ is the momentum transfer, $M_Z^{\prime}$ is the mass of the mediator, and the $(-)$ sign applies for the case of anti-neutrino scattering. 

\par As an extension, we can consider the case where the  quarks couple to $Z^{\prime}$ via a loop containing hidden sector particles~\cite{Datta:2018xty}. In this case a form factor parameterizes the quark coupling to the $Z^{\prime}$. This form factor scales as  $q^2/\Lambda^2$, where $\Lambda$ is the scale in the hidden sector associated with the mass of the mediator particle that generates the quark-hidden sector interactions. The form factor goes to unity when the quark couples directly to the $Z^{\prime}$. 
In this scenario, $g_\nu^\prime g_f^\prime$ in Eq.~(\ref{eq:SMmod}) is modified to $g_\nu^\prime g_f^\prime q^2/\Lambda^2$, where $\Lambda$ is the mass scale of the hidden sector. 

\par Motivated by the flavor content of the COHERENT stopped pion beam, in this paper we focus on  $g_e$ and $g_\mu$, associated, respectively, with electron and muon-type neutrino couplings with the $Z^\prime$. 
The up and down quark couplings to $Z^\prime$ are represented by $g_u$ and $g_d$, respectively, and we will assume that $g_u=g_d=g_\nu=g$. Under this assumption the NSI parameter defined as $\epsilon =  g_{\nu}'g_{f,v}'/\{2\sqrt{2}G_F\, (q^2 + M_{Z'}^2)\}=g^2/\{2\sqrt{2}G_F\, (q^2 + M_{Z'}^2)\}$ is positive, or $\epsilon =g^2 q^2/\{2\sqrt{2}G_F\, (q^2 + M_{Z'}^2)\Lambda^2\}$ in the hidden sector scenario. Including this parameter will have the effect of reducing the predicted number of events as it interferes destructively with the SM weak charge. On the other hand, if we set $g_u=g_d=-g_\nu=g$, the NSI parameter will become large and negative. 

\section{Data analysis}
\par
\begin{figure}[!htb]
\includegraphics[width=0.5\linewidth]{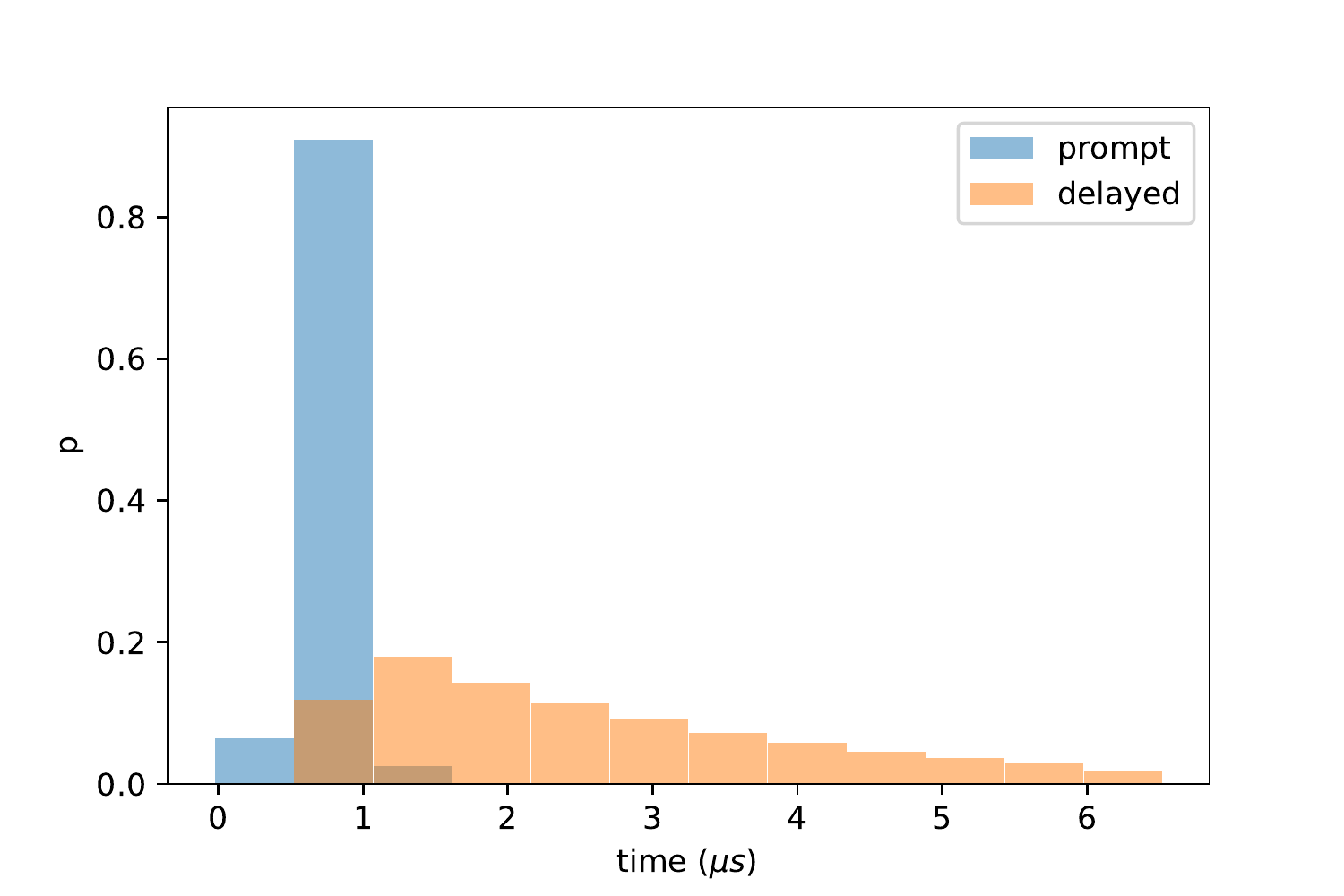}\includegraphics[width=0.5\linewidth]{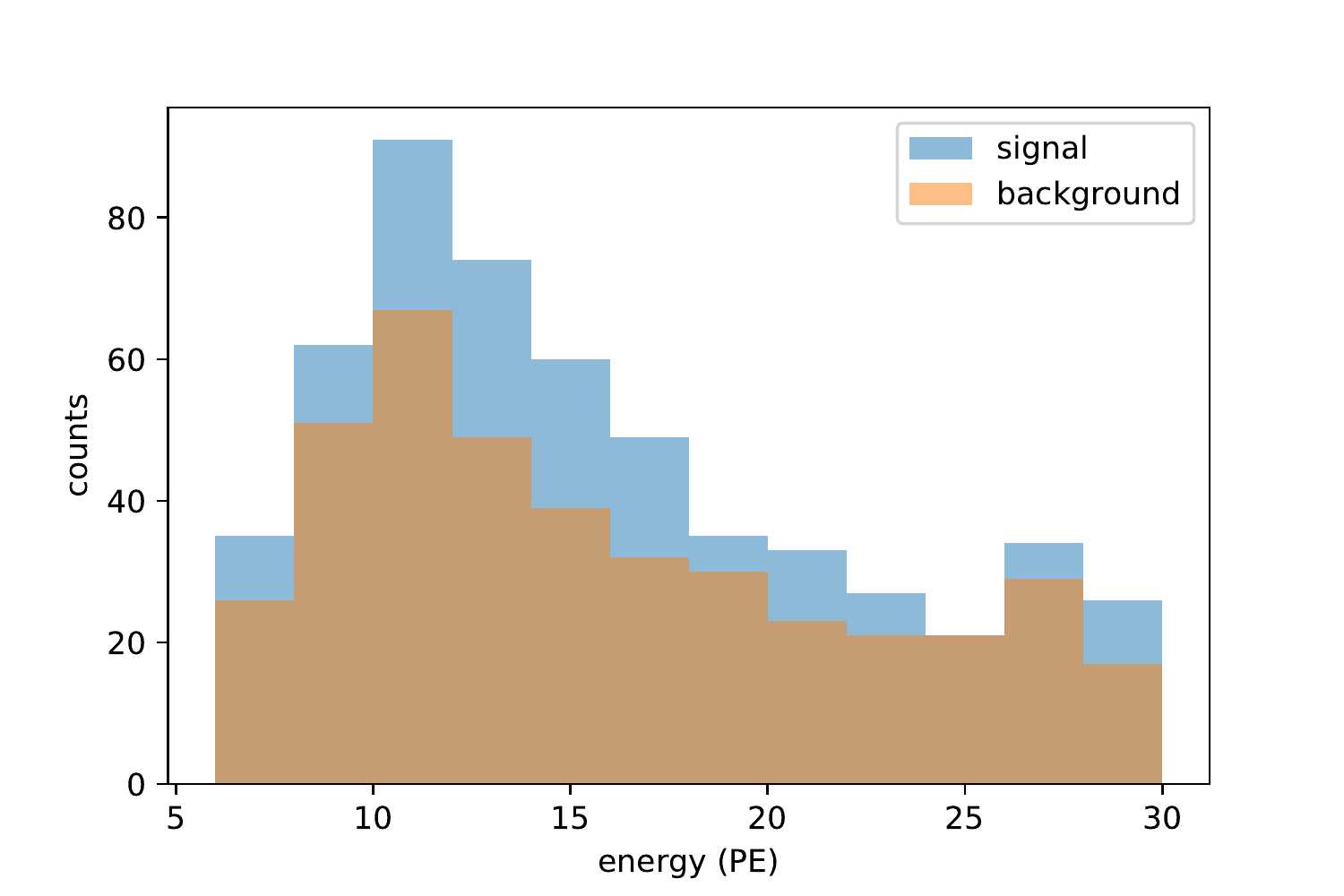}
\caption{Timing distribution for the prompt and delayed components (left) and energy distribution (right) as measured by COHERENT. The vertical axis in the time distribution gives the probability that an event is in a given time bin, while in the energy distribution it is the number of counts in that bin. In the energy distribution, the signal and background events are shown.  
} 
\label{fig:timing}
\end{figure}

 The COHERENT collaboration has published data on the energy and timing distribution of its nuclear recoil events~\cite{Akimov:2018vzs}. Figure~\ref{fig:timing} shows the energy and timing distribution of the events, with a classification of prompt and delayed events coming from the COHERENT collaboration. Note that the energy distribution is given in units of photoelectrons (PEs). Our goal is to perform a joint analysis using both the timing and energy distribution to identify possible contributions from BSM physics. To represent BSM physics, we take as parameters the couplings $g_e$, $g_\mu$, and mediator mass $M_{Z^\prime}$. To represent the uncertainty in the nuclear structure, we take $R_n$ as a parameter and use the Helm form factor model. 

\par We define $\mathcal{L}$ as the likelihood function of the data given the model parameters. To form the likelihood function, we assume that the sum of the observed nuclear recoil plus background counts, $N_{obs}(t, E)$, at time $t$ and energy $E$ follows a Poisson model with parameter  
  \begin{equation}
  \lambda(t, E)= (1+\alpha) N(t, E, \epsilon)+N_{bg}(t, E),
  \label{eq:lambda}
  \end{equation}
 where $N(t, E, \epsilon)$ is the number of neutrino-induced nuclear recoil events predicted from the theory and $N_{bg}(t, E)$ denotes the true background count. Note that by definition $N_{bg}\left(t, E\right)$ is not observed. 
 Rather, we have observed background counts, denoted by $N_{obs, bg}(t, E)$, that are proxy for the true background counts.
 We assume that given $N_{bg}\left(t, E\right)$, $N_{obs, bg}(t, E)$ follows a Poisson model with parameter $N_{bg}\left(t, E\right)$. Moreover, in the absence of any prior information about $N_{bg}(t, E)$, we use a non-informative prior on it, so that $\pi(N_{bg}(t, E))\propto 1$ for  $N_{bg}(t, E) \in [0, \infty)$.
 
 \par In addition to the counts from the signal and the background components, Eq.~(\ref{eq:lambda}) involves the uncertainty parameter $\alpha$ to account for the systematic uncertainties from flux, form factor, quenching factor, and signal acceptance uncertainties. Motivated by the results reported from COHERENT~\cite{Akimov:2017ade}, we assume this parameter follows a normal distribution with zero mean and standard deviation $\sigma_\alpha=0.28$. Defining $\vec \theta =(g_e,g_\mu, m,R_n)$ as the model parameters, the most general likelihood function for the parameters given the data is
 \be
  \mathcal{L}(\vec \theta | t, E)&\propto& \prod_{(t, E)}   \int \int \exp\{ -\lambda(t, E)\}\frac{ \{\lambda(t, E)\}^{N_{obs}(t, E)}}{  N_{obs}(t, E)! }
 \times 
  \frac{\exp (-\alpha^2/2\sigma^2_\alpha)}{\sqrt{\sigma^2_\alpha}}\nonumber \\
  &&\, \, \, \, \, \, \, \, \times \exp \{- N_{bg}(t, E)\}\frac{\{N_{bg}(t, E)\}^{N_{obs, bg}(t, E)}  }{  N_{obs, bg}(t, E)! } \,  \mathrm{d}\alpha \, \mathrm{d}N_{bg}.
  \label{eq:L}
 \ee
In comparison to previous analyses of the COHERENT data, this likelihood explicitly includes information from the timing distribution of the data. We take the bin width in energy space to be given by the bin width in the space of the number of photoelectrons, $n_e$, then convert this to recoil energy space using the relation $n_e = 1.17 (E/\textrm{keV})$. For the timing data we take the bin width directly from the COHERENT data, $\sim 0.5 \mu$s. 
We obtain the posterior probability densities for the model parameters using the multinest package~\cite{Feroz:2008xx} with flat prior distributions on the parameters. 

\par To test the robustness of our modeling in light of the steady-state (SS) background observed by COHERENT, we consider three separate scenarios for it. In the first scenario, which we refer to as background model (a), the SS background is fixed ($N_{bg}(t, E)=N_{bg, obs}$) by the reported COHERENT measurements, and the model parameters are only $(g_e,g_\mu, M_{Z^\prime},R_n)$, and we do not need to integrate over the background in Eq.~(\ref{eq:L}). 
In the second scenario, which we refer to as background model (b), we take the background shape from the COHERENT data, but allow for an overall scaling of the background to account for an uncertainty in the background normalization. 
Instead of integrating over the background for each bin, we only integrate over the total number of background counts in this scenario.
In the third scenario, which we refer to as background model (c), we take the background to be a poisson model in each bin, and integrate over the predicted number of events in each bin according to Eq.~(\ref{eq:L}). In this third scenario, in order to mitigate a bias in our reconstructed parameters, we exclude bins in which there are zero background counts. 

\par To compare any two  models, say models 0 and 1,  defined by different sets of parameters, we consider the  test statistic
$U=-2[ \log (\mathcal{L}_0)- \log \{\mathcal{L}_1(\hat \theta)\}]$. Here we define $\log (\mathcal{L}_0)$ as the log-likelihood for  model 0 in which only $R_n$ is free and other three parameters are set to zero, and $\log \{\mathcal{L}_1(\hat \theta)\}$ is the log-likelihood for the model in which at least one of the parameters in $(g_e,g_\mu,M_{Z^\prime})$ is free. For the latter model,  $\hat \theta$ denotes
the maximum likelihood estimator (MLE) of $\vec \theta$. 
The $p$-value of the test is $p=\pr(\chi^2_\eta>U_{\rm obs})$, where 
$U_{\rm obs}$ is the observed value of $U$ and $\chi^2_\eta$ is the chi-square distribution with $\eta$ degree of freedom. Here, $\eta$ is the difference between the number of estimated parameters in models 0 and 1.  A small $p$-value provides significant evidence against the null hypothesis (SM).
Applying this general procedure we can test if a model parameter (i.e., 
$g_\mu$ or $g_e$) is positive.  
We define the corresponding significance $Z$  as $\Phi^{-1}(1-p/2)$, where $\Phi^{-1}$ is the inverse cumulative distribution function of the standard normal distribution. 
Significant results can also be seen through a large value of $Z$.

\section{Results}

\par
\begin{figure}
	\includegraphics[width=0.5\linewidth]{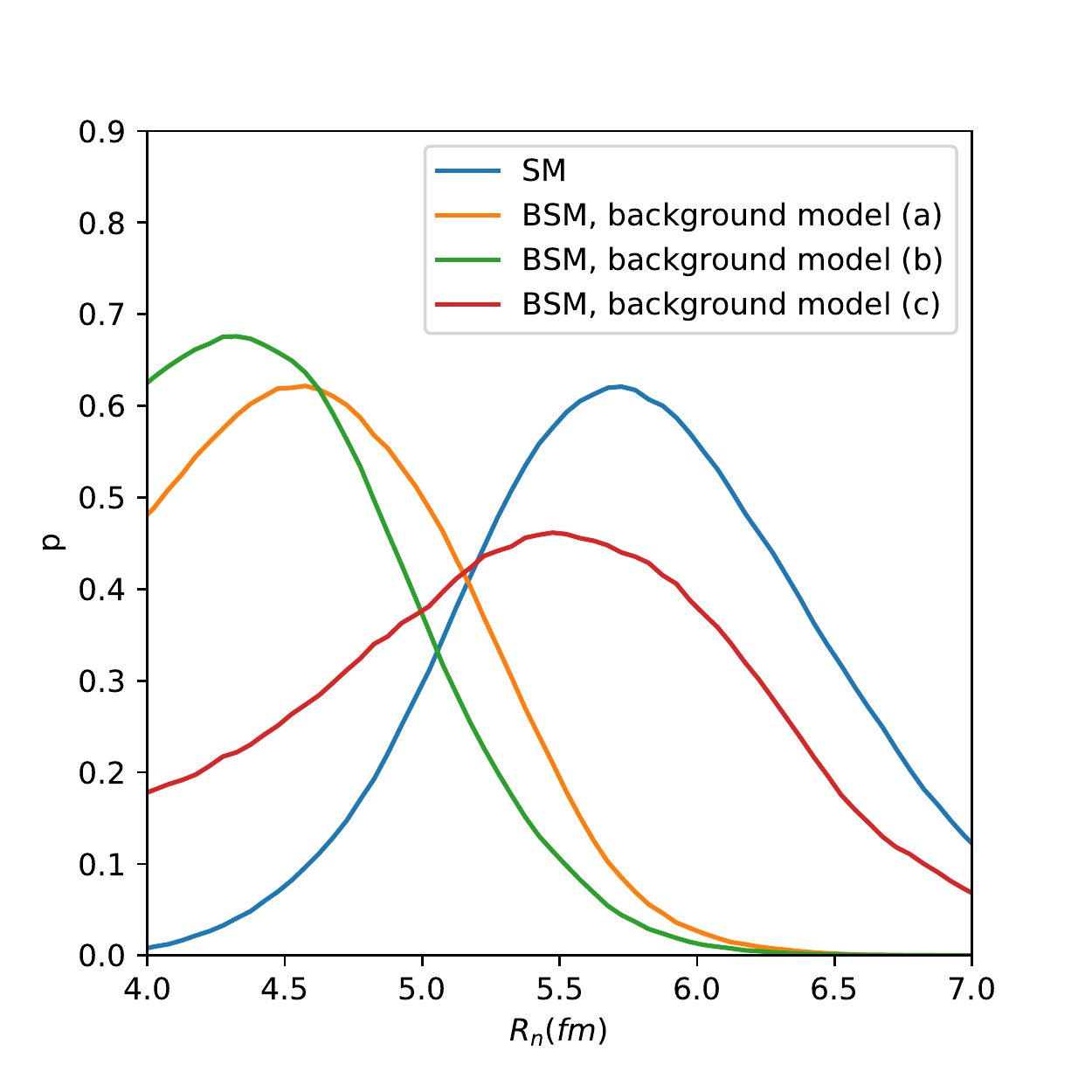}\includegraphics[width=0.5\linewidth]{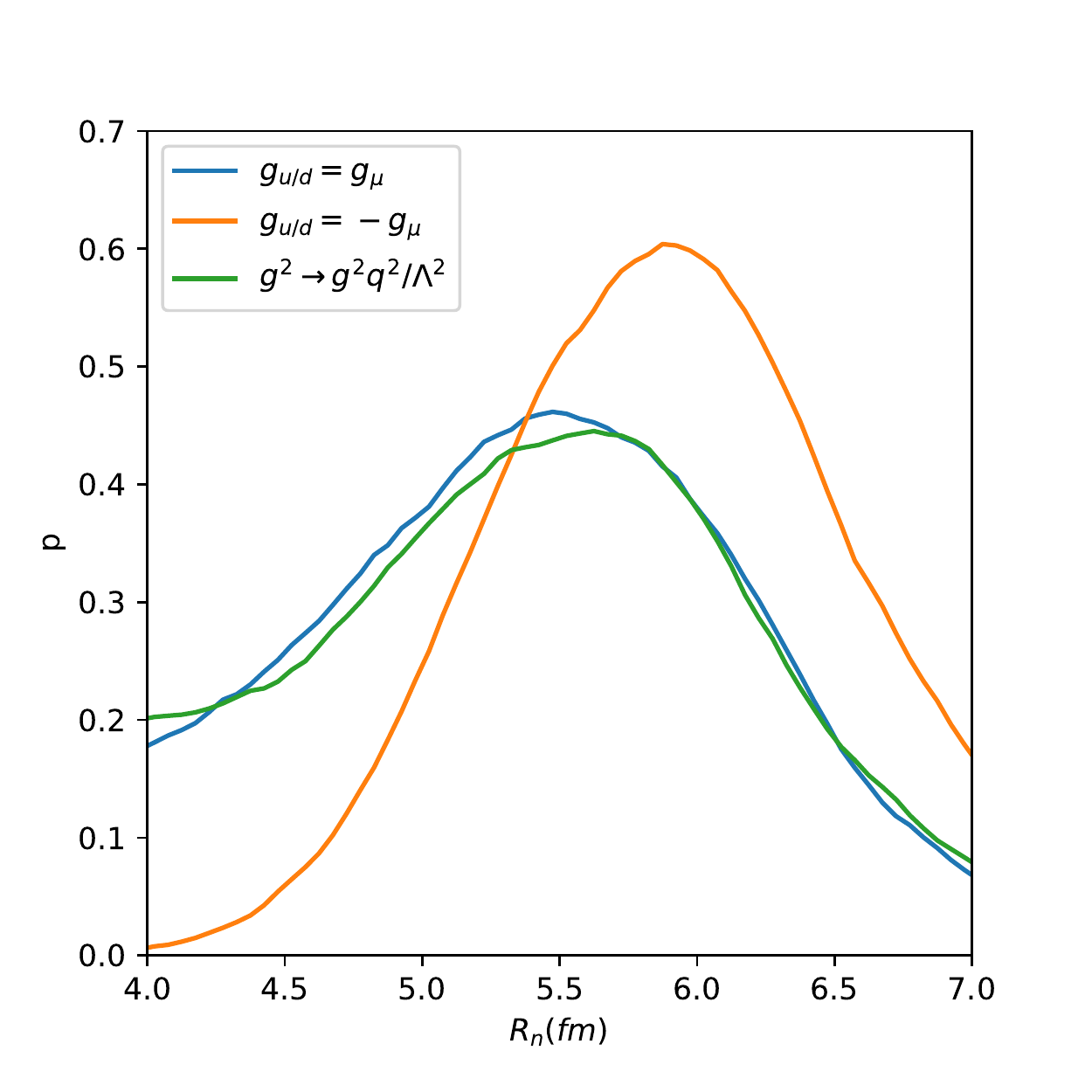}
	\caption{Posterior probability distributions for the effective size of the nucleus $R_n$.  The left panel compares different background models, and the right panel compare different scenarios for new interactions. Here we assume the bin-by-bin fluctuating background model (c) as described in the text. The (blue) SM distribution only assumes $R_n$ as a parameter, while the BSM distribution takes ($R_n$, $g^\prime$, and $M_{Z^\prime}$) as free parameters. 
	}
\label{fig:Rn}
\end{figure}

 We begin by considering how well $R_n$ is determined for each of our three different background models. The posterior probability distributions for $R_n$ are shown in the left panel of Figure~\ref{fig:Rn} for each of these background models. The value of $R_n \simeq 5.5$ fm obtained for our background model (c) is consistent with the result obtained in Ref.~\cite{Ciuffoli:2018qem}, though these authors obtained this result from background subtracted data. For background models (a) and (b), the best-fit $R_n$ is slightly lower than that from model (c), though is still statistically consistent with the result from model (c).

\par With the allowed parameter space for $R_n$ now understood, we move on to adding BSM physics in the form of light mediators. In order to simplify our analysis, for our BSM scenario we consider models with couplings such that $g_\nu=g_u=g_d=g'$. To compare the sensitivity of the data to different flavors, we either fit for $g_\mu$ and fix $g_e$, or vice versa. The key  features of our analysis are unchanged if we use a different relation among the couplings. The allowed parameter space shifts  little bit towards larger values of $R_n$ if we consider the form factor $\sim q^2/\Lambda^2$ in the quarks-$Z^\prime$ coupling.

\par
\begin{figure}
	\includegraphics[width=0.5\linewidth]{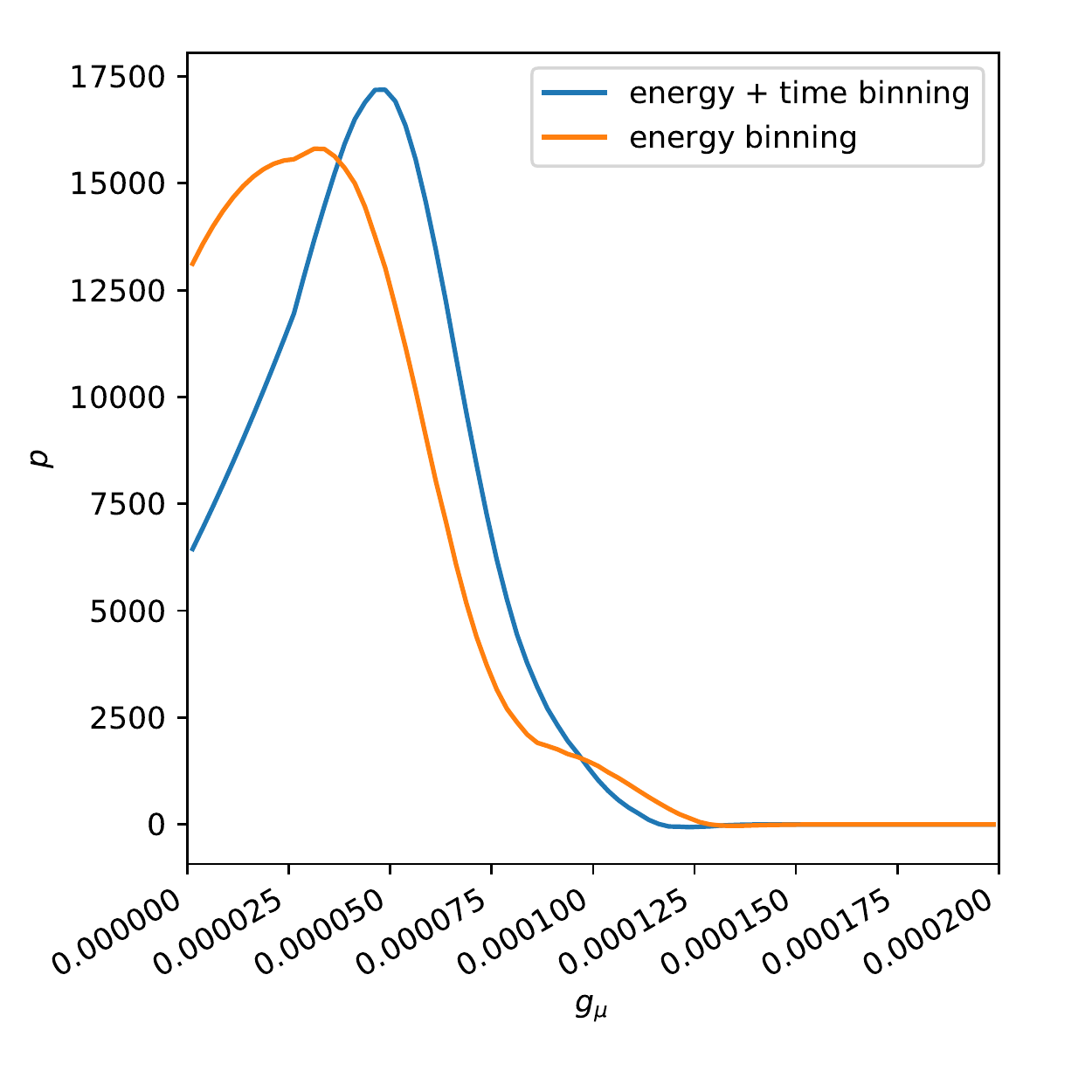}\includegraphics[width=0.5\linewidth]{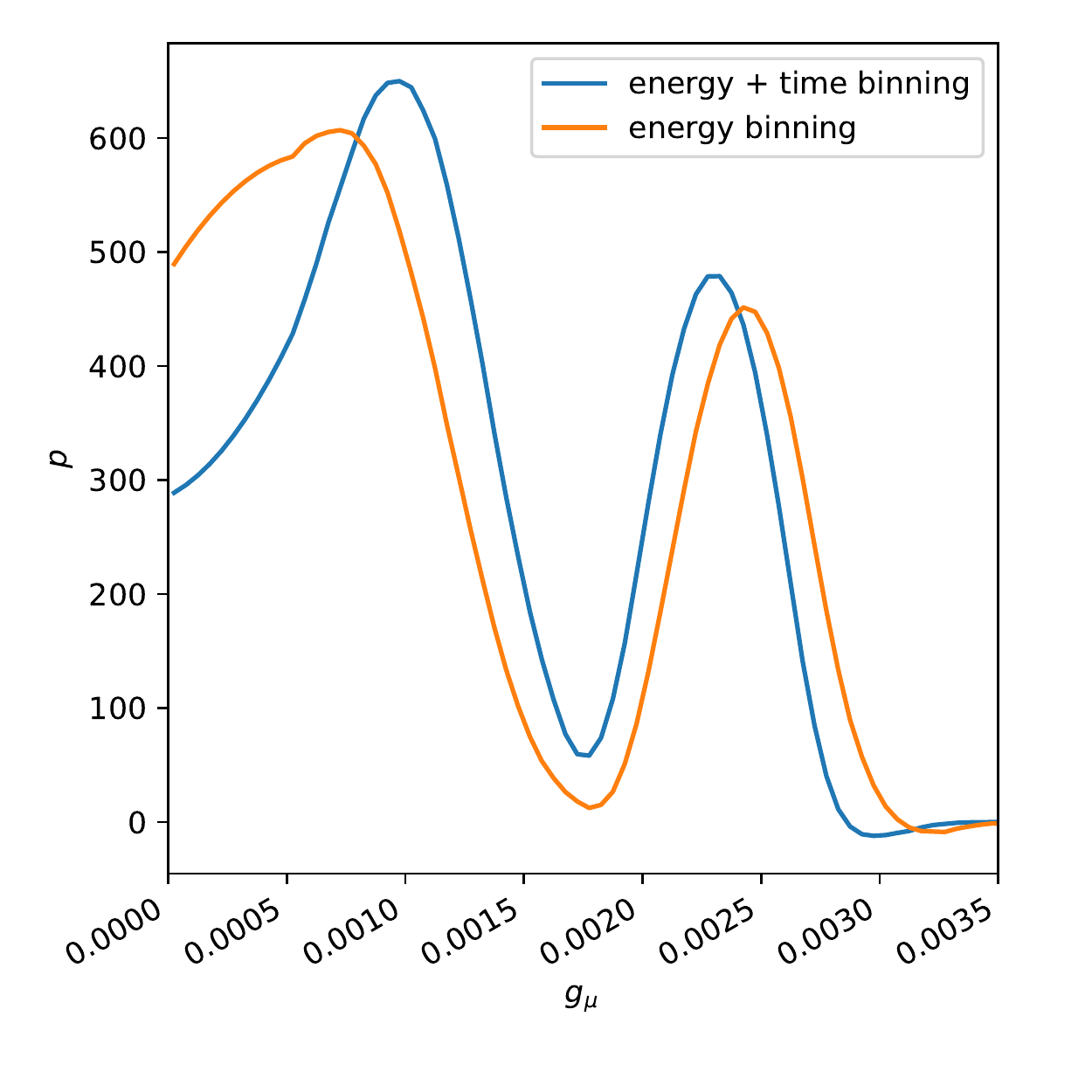}
	\includegraphics[width=0.5\linewidth]{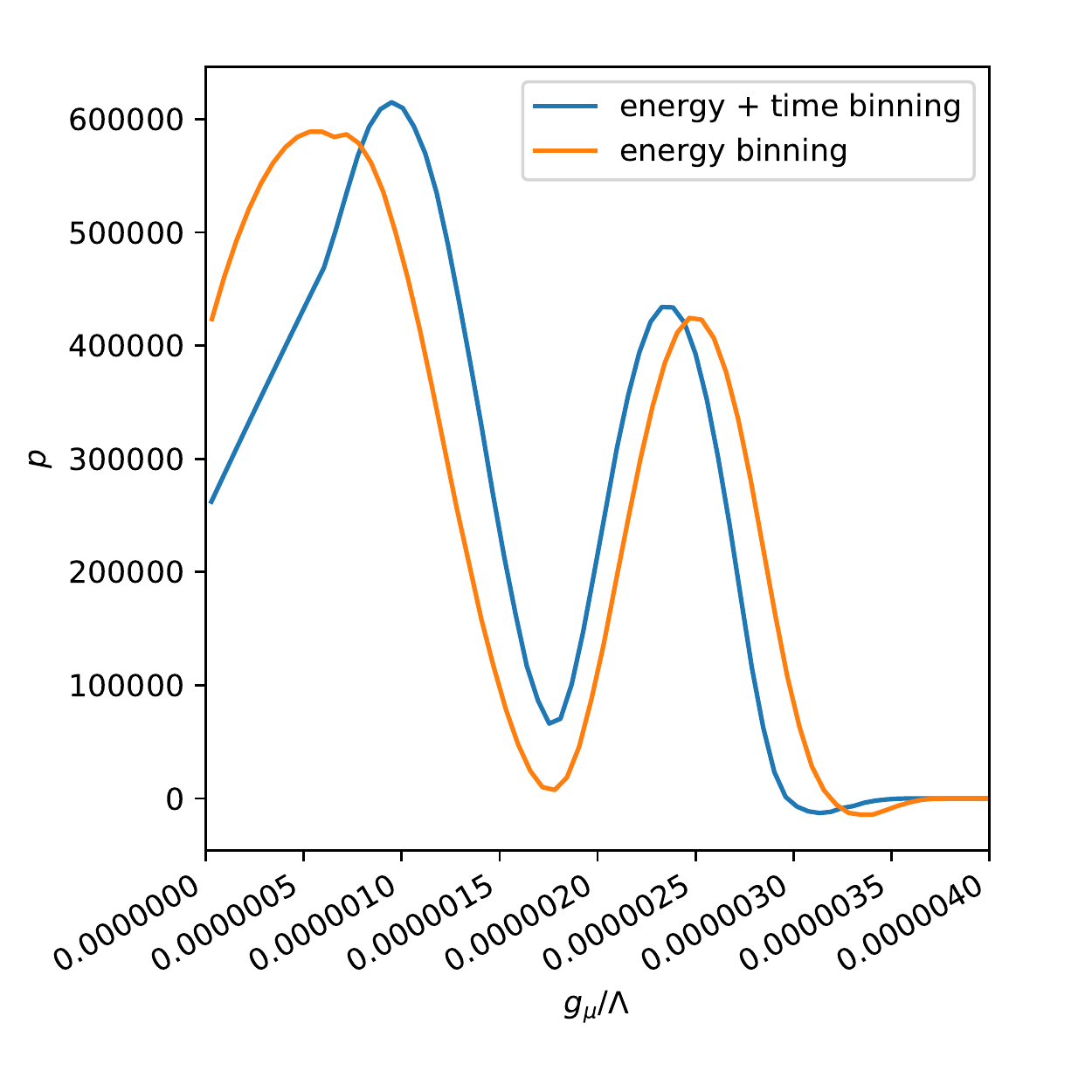}\includegraphics[width=0.5\linewidth]{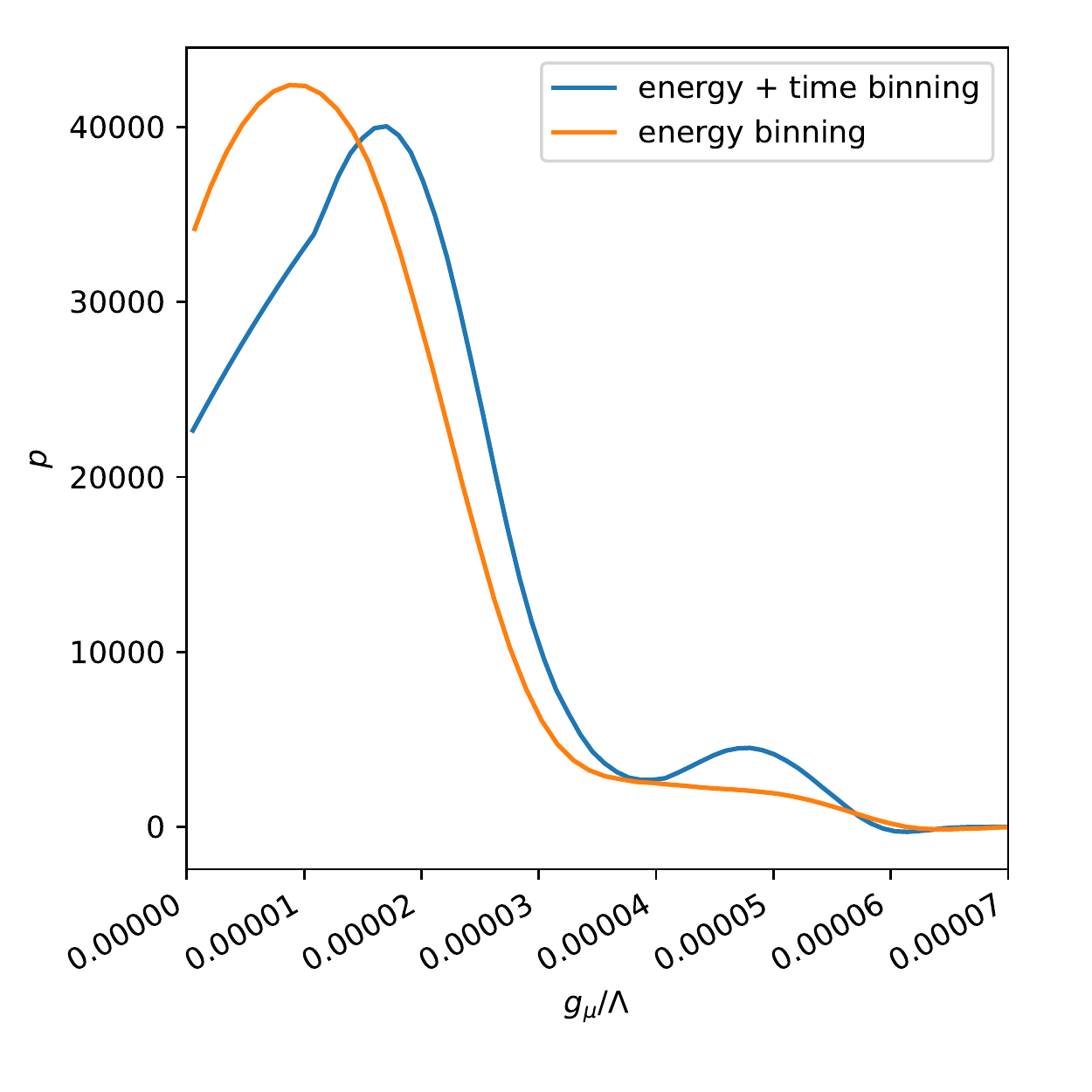}
	\caption{Posterior probability distributions for $g_\mu$ (top row) or $g_\mu/\Lambda$ (bottom row, if there is form factor), using the energy data alone (orange) and using the combined energy and timing data (blue). The left column  assumes a mediator mass of $M_{Z^\prime} = 10$ MeV,  and the right column assumes a mediator mass of $M_{Z^\prime} = 1000$ MeV. 
	}
\label{fig:gmu_2d}
\end{figure}

\begin{figure}
	\includegraphics[width=0.5\linewidth]{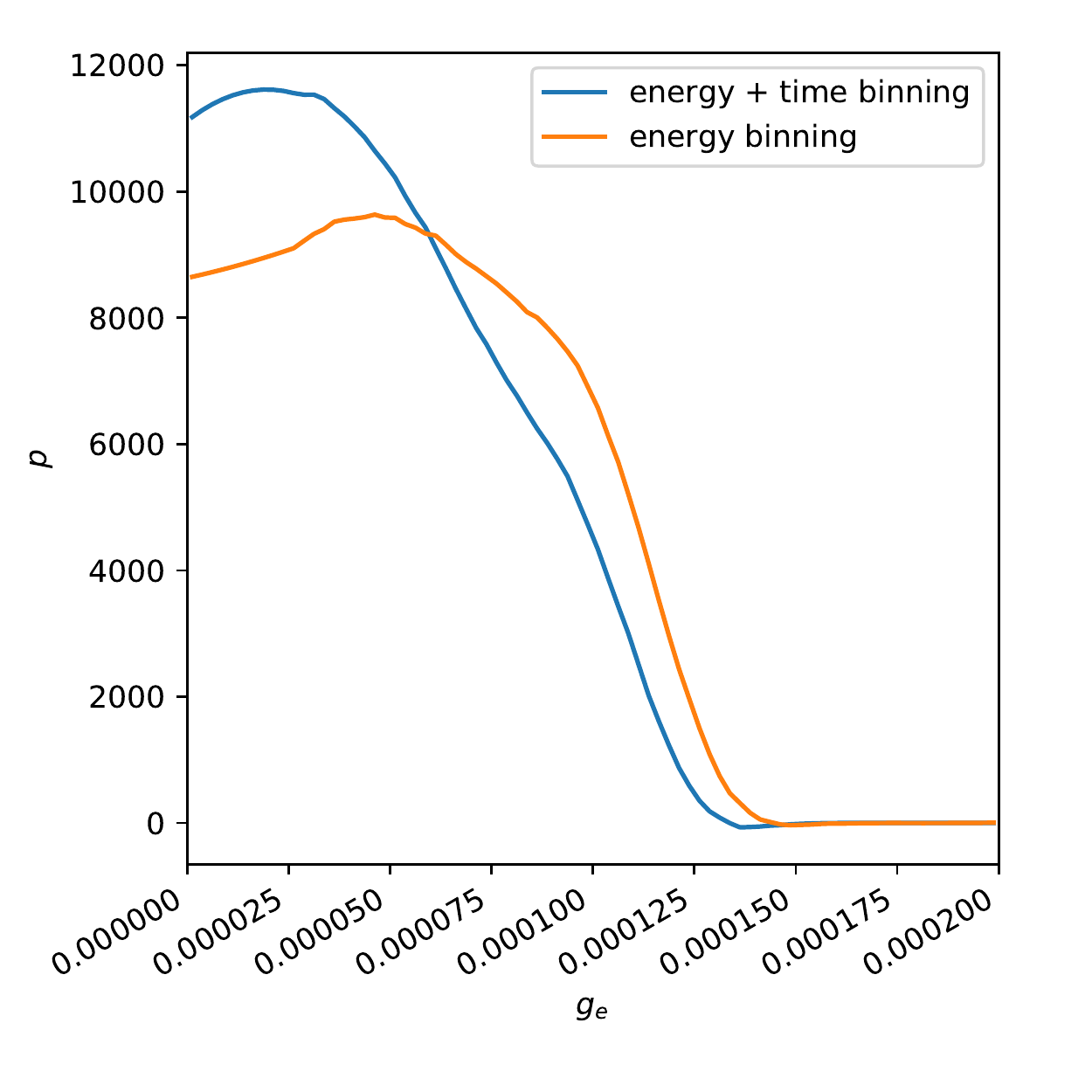}\includegraphics[width=0.5\linewidth]{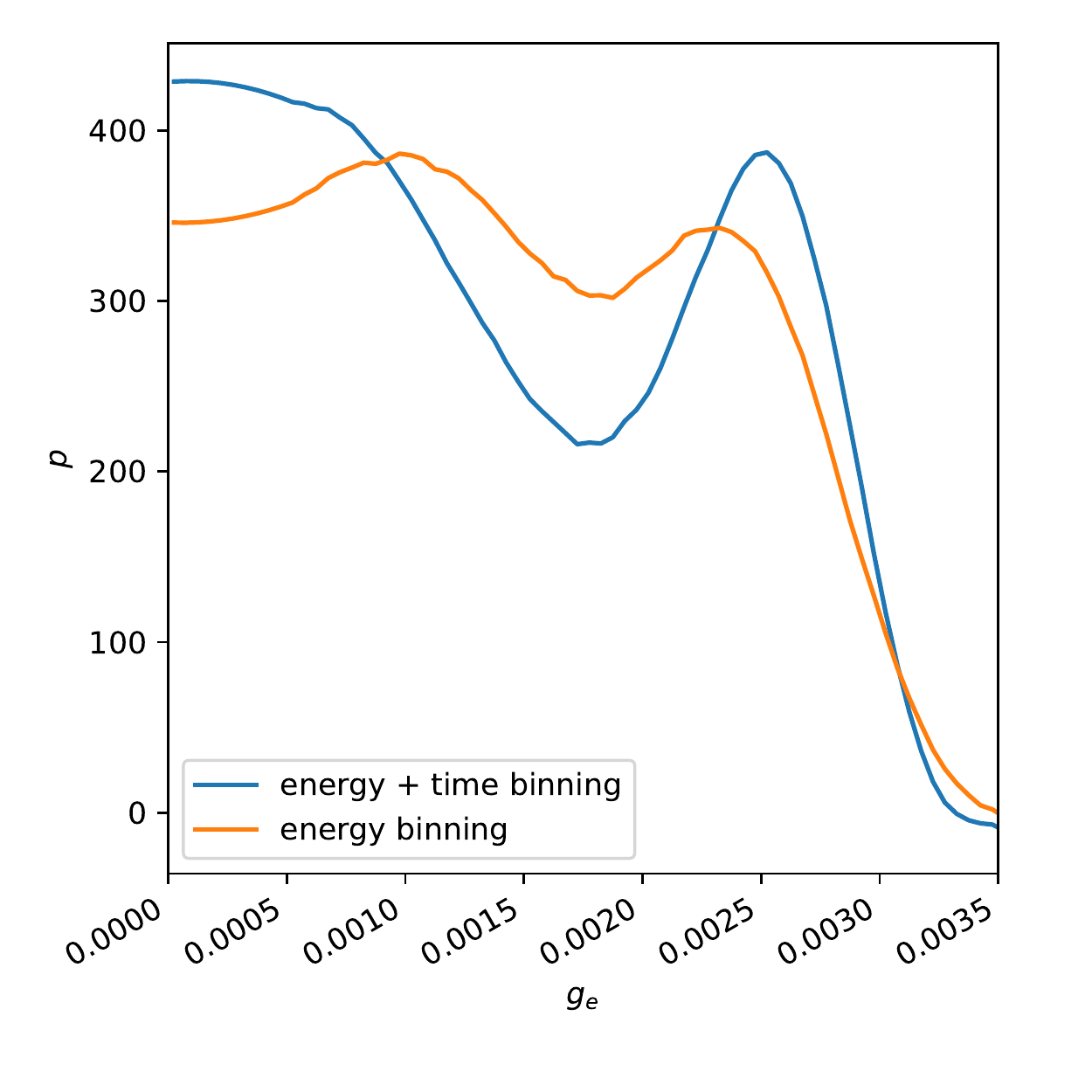}
	\caption{Same as Figure~\ref{fig:gmu_2d}, except for the coupling $g_e$.}
\label{fig:ge_2d}
\end{figure}

 In Figures~\ref{fig:gmu_2d} and~\ref{fig:ge_2d}, we show the resulting  posterior probability distributions for $g_\mu$ (assuming that $g_e = 0$) and similarly for $g_e$
(assuming that $g_\mu = 0$), for the cases of $M_{Z^\prime} = 10$ and $1000$ MeV. We show the result for background model (c), in which the background is taken to be a poisson model in each energy and time bin. The figures contain distributions using energy information alone, and distributions using both energy and timing information. For both $M_{Z^\prime} = 10$ and $1000$ MeV, the $g_\mu$ distributions are better constrained when including timing data, and deviate from the SM prediction that these couplings are zero. 

\par Though we have assumed background model (c) in Figures~\ref{fig:gmu_2d} and~\ref{fig:ge_2d}, we find that these distributions are relatively insensitive to the assumed background model. To quantify this, in Table~\ref{tab:bkgds} we show the significance $Z$ for each of the three different background assumptions. In Table~\ref{tab:bkgds} we also show the corresponding values of $Z$ for an analysis with the energy data alone. Here we find that statistical significance of the deviation from the SM is lower for all background models, as compared to the analysis that utilizes both energy and timing data. The results in Table~\ref{tab:bkgds} show that the timing data provides additional information on the flavor content of the fluxes that is not provided by the energy data alone. 
The bin-by-bin likelihood analysis that we employ is able to statistically separate the prompt and delayed distributions, with the timing information more strongly constraining the prompt $g_\mu$ component. In contrast, the $g_e$ component  only contributes to delayed neutrino recoil spectrum, and is less well constrained when adding in the timing information. 

\begin{table}
\begin{tabular}{|c|c|c|c|}
\hline 
Mediator mass, $M_{Z^\prime}$ (MeV) & Fixed (model (a)) & Fixed shape (model (b)) & Varying (model (c))\tabularnewline
\hline 
free & $1.4(0.7)$ & $0.9(0.6)$ & $1.1(0.6)$\tabularnewline
\hline 
$10$ & $1.9(1.2)$ & $1.4(1.1)$ & $1.6(1.0)$\tabularnewline
\hline 
$100$ & $1.9(1.1)$ & $1.4(1.1)$ & $1.6(1.1)$\tabularnewline
\hline 
$1000$ & $1.9(1.2)$ & $1.4(1.1)$ & $1.6(1.1)$\tabularnewline
\hline 
\end{tabular}

\caption{Significance in the likelihood ratio test for different assumed background models, as defined in the text. The first column gives the mass of the mediator that is assumed, with the first row representing the case in which the mediator mass is a free parameter. In all cases we take $g_{e}=0$ and $g_{\mu}\protect\neq0$. Each primary entry is obtained using the combined energy and timing data, and the numbers in parenthesis use just the energy data alone.
\label{tab:bkgds}
}
\end{table}

\begin{figure}
	\includegraphics{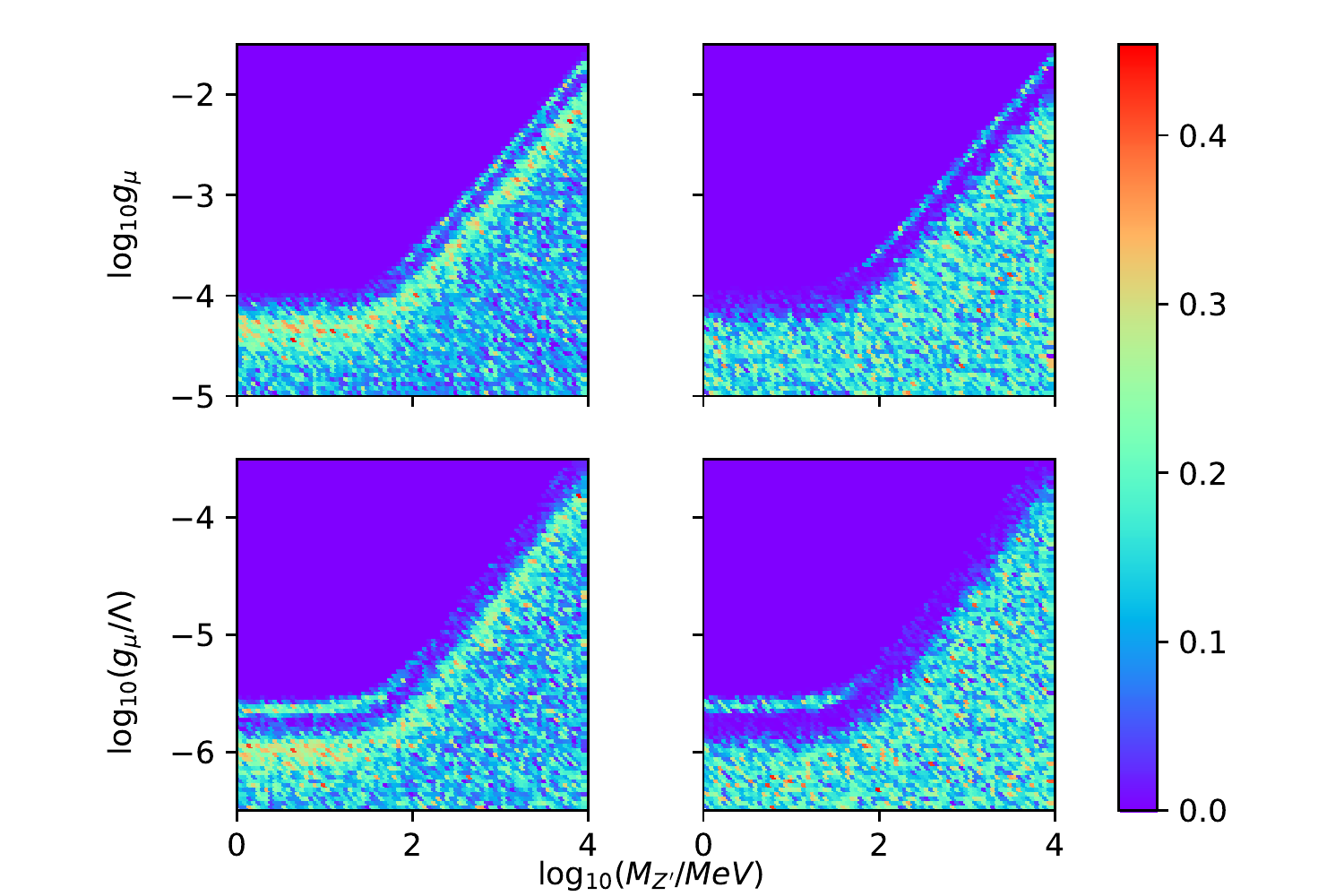}
	\caption{Heat map of the probability density in $\log_{10}(M_{Z^\prime})$ vs $\log_{10}g_\mu$ (top row) and 
	$\log_{10}(M_{Z^\prime})$ vs $\log_{10}(g_\mu/\Lambda)$ (bottom row, when there exists a form factor $q^2/\Lambda^2$) parameter space using energy and timing data (left) and energy data alone (right). Here we assume background model (c). 
	}
\label{fig:heat}
\end{figure}

Figure~\ref{fig:heat} shows the probability density in $\log_{10}(M_{Z^\prime})$ vs $\log_{10} (g_\mu$) or $\log_{10}(g_\mu/\Lambda)$ space. From the figure we can see for mediator mass from 1 MeV to 1 GeV the energy and timing data imply a best-fit deviation from the SM, while the energy data alone is more consistent with the SM. The shape of the boundary in both plots can be understood as follows: in the small mediator mass region, $q^2 \gg M_{Z^\prime}^2$, the NSI parameter is independent of the small mediator mass, while in the large mediator mass region, $M_{Z^\prime}^2 \gg q^2$, the NSI parameter depends on $g^2/M_{Z^\prime}^2$, thus, in log space the slope is about $1$. The isolated island at large mediator mass region is because the global degeneracy for the weak charge across all energy bins (since the NSI parameter is independent of energy). On the contrary, if a hidden sector is introduced to generate a form factor $\sim q^2/\Lambda^2$,  the NSI parameter becomes independent of energy in the smaller mass region and consequently, the degeneracy appears in the smaller mass region.

\par The above results favoring $g \ne 0$ are independent of the uncertainty in the assumed value for $R_n$, since we have taken $R_n$ as a parameter in our analysis. It is however interesting to determine the manner in which $R_n$ is degenerate with the BSM model. As an example we can compare to the case in which $g_u=g_d=-g_\nu=g$, as in this case there is destructive interference with the SM weak charge. The resulting $R_N$ distribution with this assumption is shown in the right-hand panel of Fig.~\ref{fig:Rn}. Here we see that the degeneracy is such that larger $R_N$ is now more favored, though the results are statistically consistent with the assumption of $g_u=g_d=g_\nu=g$. We stress again that this distribution of $R_n$ is likely maximally conservative, and that input from the nuclear model of CsI will help break this degeneracy. 

\begin{table}

\begin{adjustwidth}{-4em}{-4em}
\centering
\begin{tabular}{|c|c|c|c|}
\hline 
$M_{Z^\prime}$ (MeV) & 10 & 100 & 1000\tabularnewline
\hline 
$g_\mu$ & $[1.87,6.65]\times 10^{-5}$ & $[0.41,1.47]\times 10^{-4}\oplus[2.47, 2.66]\times 10^{-4}$ & $[0.48,1.32]\times 10^{-3}\oplus[2.17, 2.47]\times 10^{-3}$\tabularnewline
\hline 
$g_e$ & $[0,6.12]\times 10^{-5}$ &  $[0,1.53]\times 10^{-4}\oplus[2.53, 2.84]\times 10^{-4}$ & $[0,1.22]\times 10^{-3}\oplus[2.22, 2.77]\times 10^{-3}$\tabularnewline
\hline 
\end{tabular}
\end{adjustwidth}

\caption{The $1\sigma$ constraining windows on $g_\mu$ and $g_e$ using energy plus timing information. The results are shown using background model (c).
\label{tab:window}
}
\end{table}

\par To quantify the constraints on the NSI parameters in Fig.~\ref{fig:gmu_2d} and Fig.~\ref{fig:ge_2d}, we show in Table~\ref{tab:window} the $1\sigma$ window of $g_\mu$ and $g_e$ assuming 10, 100, and 1000 MeV mediator masses using our results from the energy and timing analysis. Note that this range is consistent with general constraints on the couplings and masses of light mediators~\cite{Bauer:2018onh}.

\par Constraints on the NSI parameters also come from neutrino oscillation experiments. To obtain constraints from neutrino oscillation experiments, the NSI parameters are expressed as $\epsilon=g^2/\{2\sqrt{2}G_F\, (M_{Z'}^2)\}$. Since the neutrino undergoes forward scattering off matter, it is independent of $q^2$. The existing constraints from the neutrino oscillation experiments are consistent with our result, within the $\sim 2\sigma$ range~\cite{Coloma:2017egw}. For example, the allowed range for a heavy mediator for $\epsilon^u_{\mu\mu}$ is $[0, 0.005]$ (our model only has positive $\epsilon$). This converts to $g = [0, 0.0013]$ at $M_{Z^\prime} = 1~\mathrm{GeV}$, which covers the $1\sigma$ region as shown in Fig.~\ref{fig:gmu_2d}.

\par There also exists so-called a LMA-Dark solution, which is characterized by larger NSI couplings and is not strongly constrained by oscillation experiments. Ref.~\cite{Denton:2018xmq} discusses this scenario specifically in the context of the COHERENT data. Since the SM is not nested within the LMA-Dark solution, to establish significance in this case we must appeal to the Bayes factor. We find that the Bayes factor is $1.12~(0.85)$ with (without) timing information, which means this kind of NSI is also slightly preferred over SM if we include the timing data. 

\section{Discussion and conclusion}
\par In this paper we have performed a fit to the energy and timing distribution of nuclear recoil events from the COHERENT data. A combined timing and energy analysis provides information on the flavor content of the neutrino flux beyond what is obtained with the energy data alone. We have shown that including the information in both the energy and timing distributions of the COHERENT data, there is a $\sim 2\sigma$ deviation between the best-fitting model and the SM prediction. Light mediators in the mass range $\sim 10-1000$ MeV are able to provide a good fit to the data. Though our primary analysis is performed within the context of models with $g_\nu=g_u=g_d=g'$ for light mediators, in the appropriate limits of large mediator mass, $\gtrsim 1$ GeV, this analysis is equivalent to a heavy mediator NSI analysis that has been performed by the COHERENT collaboration and by previous authors. We analyzed a BSM scenario where  an additional factor $\sim q^2/\Lambda^2$ exists in the  quarks-$Z^\prime$ coupling arising from the hidden sector interactions. We showed the $1\sigma$ allowed window of $g$ in our model. Our result is consistent with dark photon searching experiments as well as oscillation experiments. We also showed the energy plus timing information can be used to constrain LMA-Dark solution of neutrino oscillation, and that LMA-Dark solution is also prefered than SM with Bayes factor $1.12$.

\par The analysis that we have performed in this paper highlights that there is additional statistical information in an event-by-event analysis of COHERENT nuclear recoil events. Given information on individual nuclear recoil events, our analysis may be extended in the future by considering an unbinned limit to the likelihood, and may include appropriate uncertainty on the energy and time of an event. Additional BSM physics may then ultimately be tested, e.g. in the form of sterile neutrinos or neutrino magnetic moment.  

\par Future data from COHERENT, as well as from reactor-based experiments, will better clarify the results that we have obtained. An important expected near-term result is COHERENT data using different nuclear targets~\cite{Akimov:2018ghi}. In addition to increasing the exposure, different nuclear targets will help further decrease the uncertainties that arise due to the nuclear structure.

\section*{ACKNOWLEDGMENTS}
BD and LES acknowledge support from DOE Grant de-sc0010813. We acknowledge support from COS-STRP (TAMU), and would like to thank Grayson Rich and Kate Scholberg for discussions on this paper. 

\bibliography{main}
\end{document}